\documentclass[a4paper,conference]{IEEEtran}
\usepackage{cite}
\usepackage[british]{babel}
\usepackage[binary-units]{siunitx}
\usepackage{calc}
\usepackage{graphicx}
\usepackage{textcomp}
\usepackage{xcolor}
\usepackage[obeyspaces,hyphens]{url}
\newcommand{\URL}[1]{$\langle$\url{#1}$\rangle$}
\newcommand*{\TakeFourierOrnament}[1]{{%
\fontencoding{U}\fontfamily{futs}\selectfont\char#1}}
\newcommand*{\danger}{\TakeFourierOrnament{66}}
\usepackage{wasysym}
\IEEEoverridecommandlockouts
\IEEEpubid{\makebox[\columnwidth]{000-0-0000-0000-0/00/\$00.00~\copyright
2019 IEEE \hfill}
\hspace{\columnsep}\makebox[\columnwidth]{ }}
\begin{document}
\title{(``Oops! Had the silly thing in reverse'')---Optical injection attacks
in through LED status indicators}
\author{\IEEEauthorblockN{Joe Loughry,~\IEEEmembership{Senior Member,~IEEE}}
\IEEEauthorblockA{University of Denver \\
Colorado, USA \\
Email: joe.loughry@cs.du.edu, or joe@netoir.com}\thanks{Phase 0 results of this
research were orally presented at EMC Europe 2018, in Amsterdam, the
Netherlands (\emph{International Symposium and Exhibition on Electromagnetic
Compatibility}), 27--30 August 2018.}}
\date{28 February 2019}
\maketitle
\begin{abstract}
It is possible to attack a computer remotely through the front panel LEDs.
Following on previous results that showed information leakage at optical
wavelengths, now it seems practicable to inject information into a system as
well. It is shown to be definitely feasible under realistic conditions (by
infosec standards) of target system compromise; experimental results suggest it
further may be possible, through a slightly different mechanism, even under
high security conditions that put extremely difficult constraints on the
attacker. The problem is of recent origin; it could not have occurred before a
confluence of unrelated technological developments made it possible.
Arduino-type microcontrollers are involved; this is an Internet of Things (IoT)
vulnerability. Unlike some previous findings, the vulnerability here is
moderate---at present---because it takes the infosec form of a classical covert
channel. However, the architecture of several popular families of
microcontrollers suggests that a Rowhammer-like directed energy optical attack
that requires no malware might be possible. Phase I experiments yielded
surprising and encouraging results; a covert channel is definitely practicable
without exotic hardware, bandwidth approaching a Mbit/s, and the majority of
discrete LEDs tested were found to be reversible on GPIO pins. Phase II
experiments, not yet funded, will try to open the door remotely.
\end{abstract}
\section{The Nature of the Vulnerability}
Light emitting diodes (LEDs) are reversible; when illuminated by an outside
source, they act like tiny solar cells and produce an electric current. They
can be used as photodiode light sensors. The fact has been known since at least
the nineteen-seventies and is occasionally useful \cite{Mims1973b, Mims1977a,
Mims1979a, MERL2003a, Karadaglic2007a, Mims2014a}. Modern
microcontrollers---those small microprocessors used for all kinds of things
that aren't generally thought of as `computers'---are highly integrated
systems-on-a-chip (SoC) that directly incorporate most of the peripheral
devices---such as memory, network interfaces, and digital-to-analogue
converters---that used to be separate chips interfaced directly with the
address and data buses of the central processing unit (CPU). Consequently, new
microcontrollers tend to expose instead a large number of general purpose
input/output (GPIO) pins to the outside world for the purpose of reading
buttons and switches, turning on LEDs, controlling small motors, or connecting
to a speaker. GPIO pins are highly adaptable; any GPIO can be a digital input
or output, and some of them can be analogue inputs or outputs. They can be
pulse width modulation (PWM) sources, or serial channel clock and data signals.
The behaviour of any given GPIO pin at any particular time is controlled by
software running on the CPU \cite{Thornton2019}.

One other thing about GPIO pins: these days they have programmable pull-up or
pull-down resistors---a great convenience to hardware
designers.\footnote{Pull-up resistors---and in some cases pull-down
resistors---are needed for reading switches or buttons reliably in the real
world, and sometimes when resources on the circuit board are shared.} In the
old days, it was necessary to install components on the board, taking space and
driving up costs. Now the designer can do all it in software. (And software
can be changed.)
\begin{figure}[ht]
  \centering
  \includegraphics[width=\columnwidth]{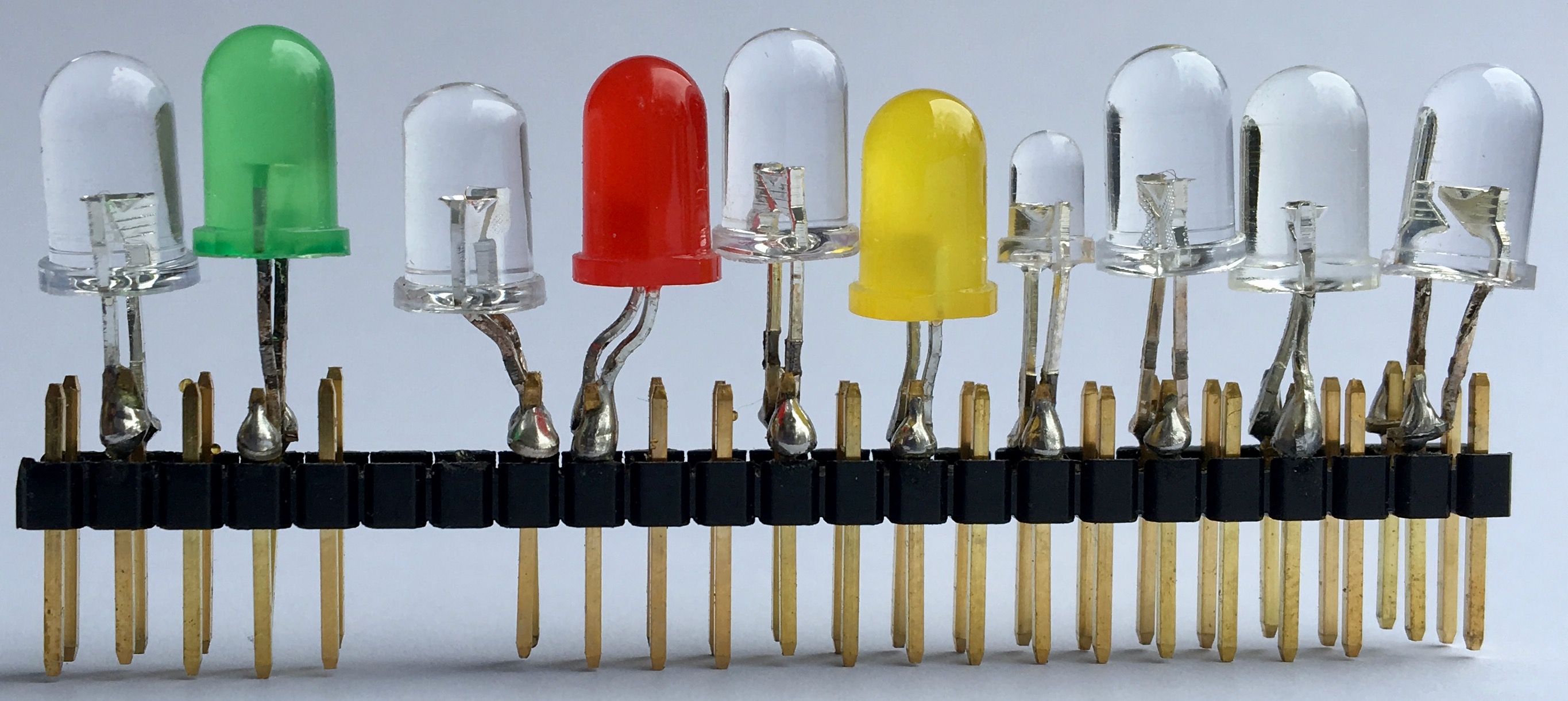}
  \caption{A collection of light emitting diodes. Not only do some LEDs leak
  information, but information can leak in through them. Of the ten devices
  shown here, only three (the green diffuse, the red diffuse, and the smallest
  one---fourth LED from the right) failed to exhibit any exploitable response.}
  \label{figure:test_subjects}
\end{figure}
Hardware designers have lots of GPIO pins to use, and use them for all sorts of
things, including LEDs. The third and final piece of the puzzle (see Table
\ref{table:timeline}) was the introduction of ultra high-brightness LEDs
\cite{Stringfellow1997}. These are highly efficient, so obviate the need for
external current-buffering transistors. They also---although the phenomenon has
not really been studied by anyone---seemingly have at least as high a
responsivity to external light sources as older discrete LEDs did (see Figure
\ref{figure:test_subjects}). The combination is deadly. Reprogrammable GPIO
pins connected to LEDs are an unlocked door through which hackers can send
information in both directions.

This is a preliminary report of work in progress. It has been recently reduced
to practice and yielded some interesting results. Some practicable attacks have
been developed.
\begin{table}[ht]
  \centering
  \includegraphics[width=\columnwidth]{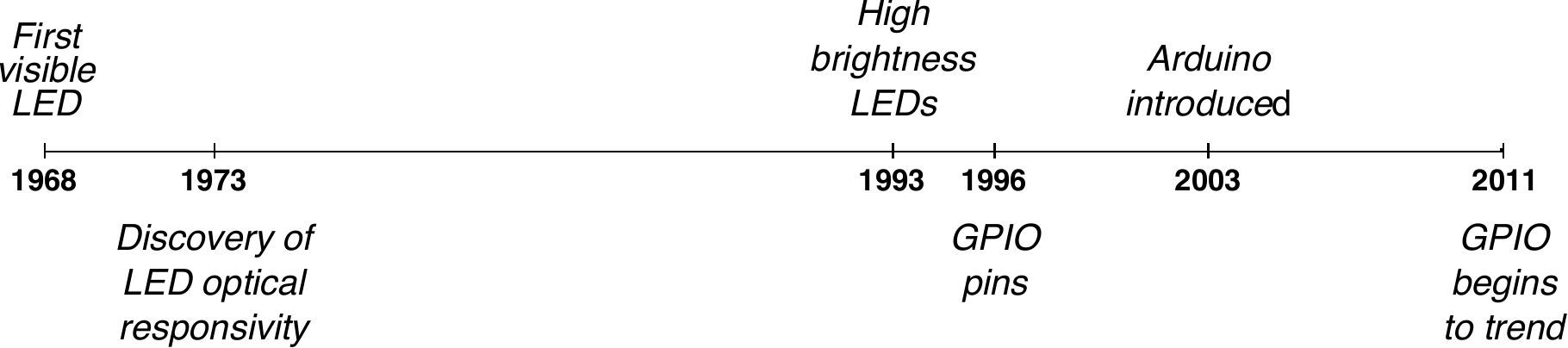}
  \caption{A confluence of several technological developments made
    the possibility of reversing LEDs into a vulnerability now.}
  \label{table:timeline}
  \vspace{-5mm}
\end{table}
\section{Optical TEMPEST}
It has long been known that little blinky lights leak information
\cite{Loughry2002a}.
Depending on the nature of the leakage, information leakage through optical
emanations may or may not be a covert channel \cite{Lampson1973}. By the
classical definition, a covert channel comprises a pair of communicating
processes on the CPU, but in many cases, the source of compromising optical
emanations is an accidental function of the hardware, not a design function of
some software running unbeknownst to the data owner.

Most published research related to compromising optical emanations---`optical
TEMPEST' radiation---concerns information flow out of the computer system; only
a few papers in the literature have anything to say about information flow
inwards
\cite{Domburg2006, Rieback2006, Nassi2017a}.
\begin{figure}[ht]
  \centering
  \includegraphics[height=1.25in]{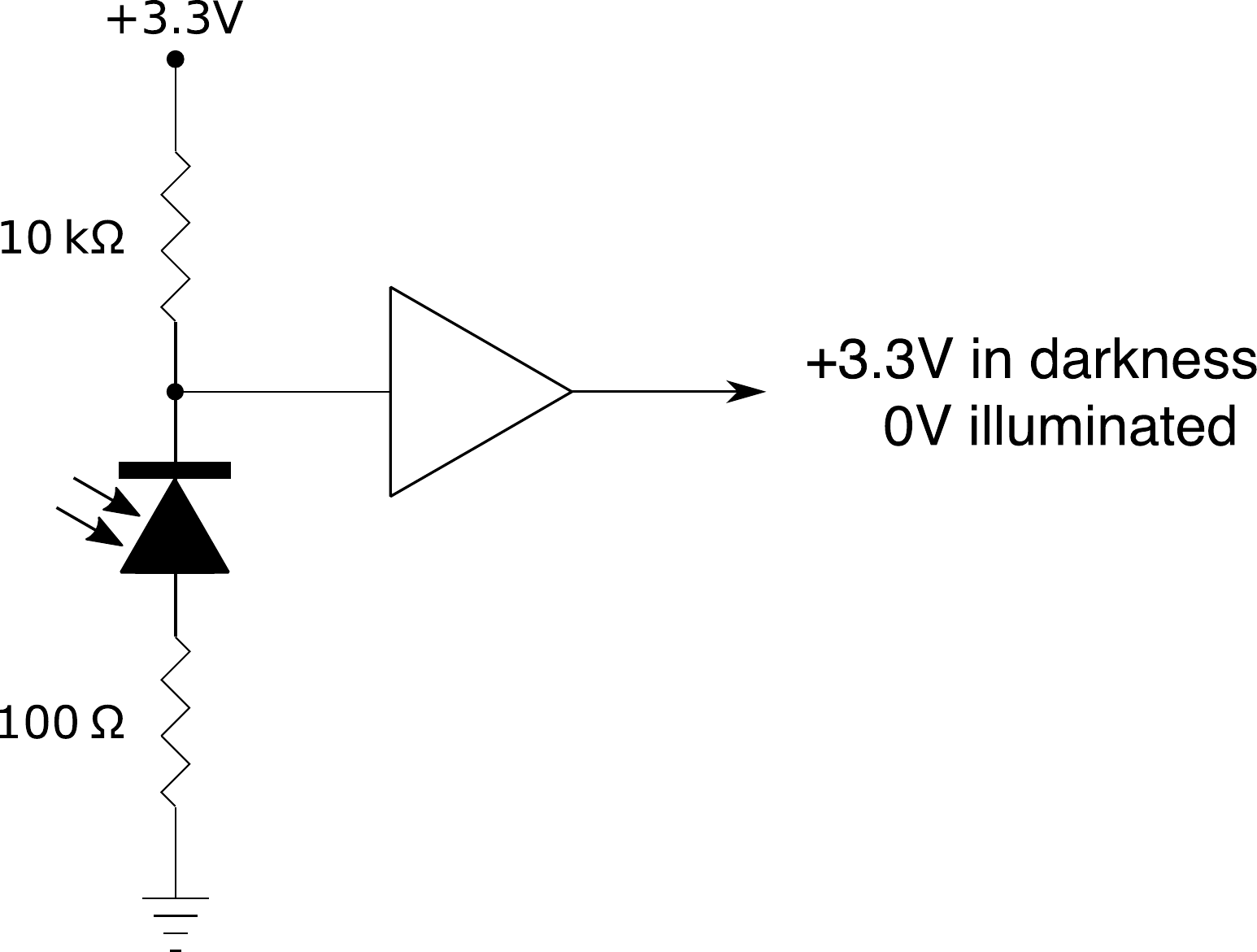}
  \caption{Conventional (reverse biased) photodiode in photoconductive mode of
    operation.}
  \label{figure:conventional_photodiode}
\end{figure}
\section{Reversing LEDs}
Inadvertent photosensitivity of electronics is a well-known phenomenon. In
1952, the installation of the first production IBM 701 mainframe computer was
disrupted when the bright
flashbulbs of news photographers, invited for the occasion, erased the Williams
tube memory of the computer \cite{Jennings1990}. Donald E.~Rosenheim, one of
the IBM engineers, said afterwards:
\begin{quote}
\emph{Suffice it to say that shortly thereafter the doors to the CRT storage
frame were made opaque to the offending wavelengths \cite{Rosenheim1983}.}
\end{quote}
Usually, the problem is mitigated by opaque chip packages; in one application
note for a Wafer-Level Chip-Scale Package (WLCSP), a backside laminate (BSL) is
an optional feature that can protect flip-chip components from being affected
by light \cite{ONsemi2018}.
In the case of eraseable programmable read-only
memory (EPROM) chips, the photoelectric effect is used to advantage by
providing a quartz
window transparent to ultraviolet (UV) radiation for erasing the memory. (The
window is covered by a sticker when memory contents are wanted.) Early
memory devices, with the top prised off to expose the silicon chip to light via
a lens, could be re-purposed as digital cameras \cite{Micron1984,
Whitehead1984, Kurz1994}. The problem has occurred in production as recently
as 2015 when a power regulator on the Raspberry Pi 2 single-board computer (SBC)
was found to be susceptible to camera flashes and laser pointers
\cite{Upton2015}.

Since first being reported by Mims (1973), the phenomenon of LED responsivity
to external radiation has been used a few times for legitimate purposes. But
until recently it was not an exploitable vulnerability, in general because the
attacker would need to move wires around inside equipment in order to have an
LED---with the polarity right way round---connected to an input. That changed
with the advent of GPIO pins and programmable pull-up/pull-down resistors. Now
an attacker with the capability only to alter software running on the device
could simply re-assign any pin the designer meant for an output to be an input;
further, depending on circumstances, the attacker might even be able to
compensate for polarity mismatch.
\begin{figure}[ht]
  \centering
  \hfill
  \begin{minipage}[b]{0.27\columnwidth}
    \centering
    \includegraphics[width=\textwidth]{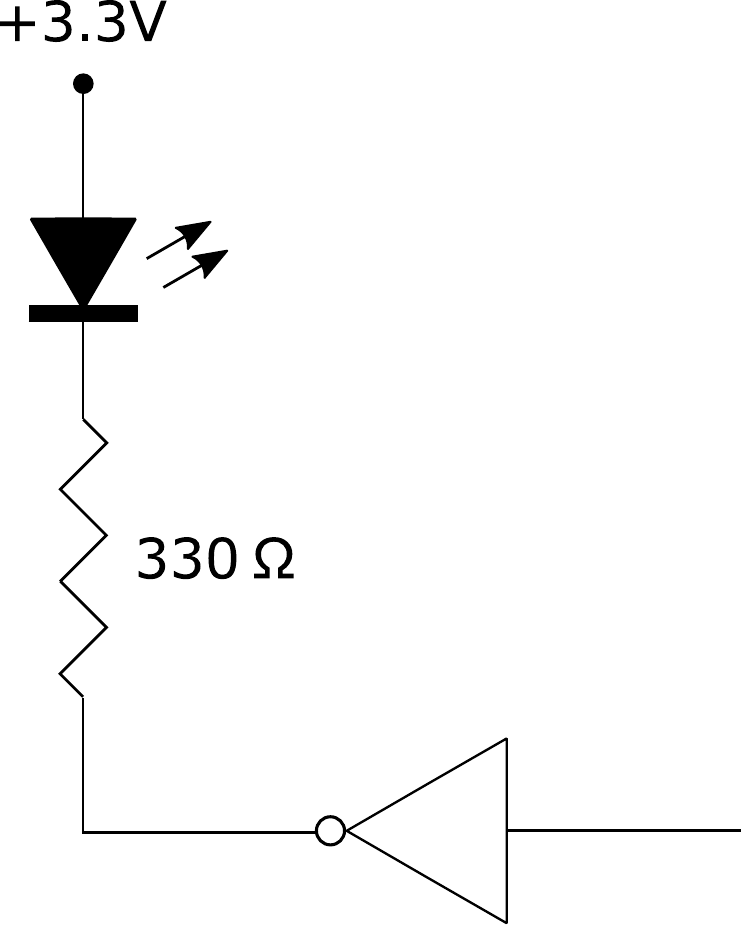} \\
    (a)
  \end{minipage}
  \hfill
  \begin{minipage}[b]{0.3\columnwidth}
    \centering
    \includegraphics[width=\textwidth]{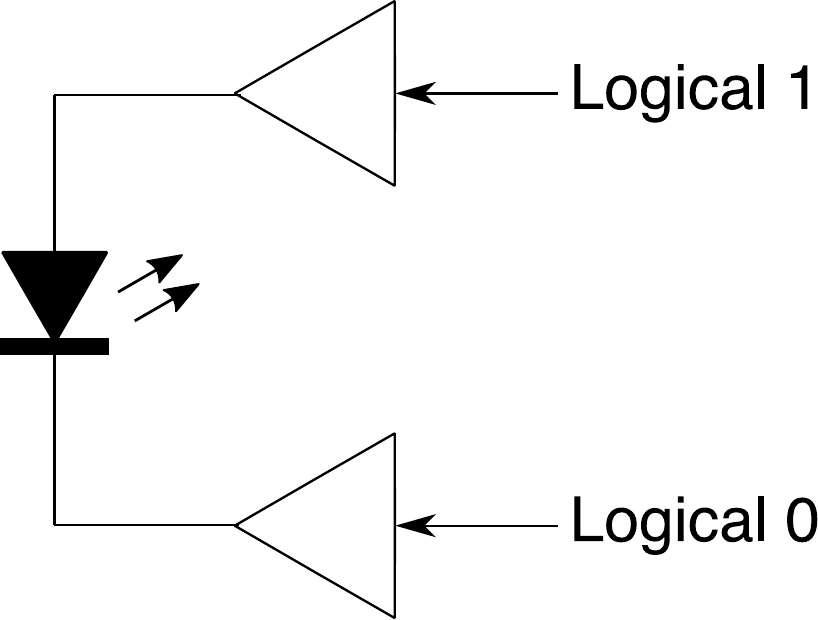} \\
    (b)
  \end{minipage}
  \hfill\strut
  \caption{Reasonable and unreasonable---but sometimes convenient---ways of
    driving LED indicators on modern microcontrollers.}
  \label{figure:LED_drivers}
\end{figure}

\subsection{GPIO Pins}
It is worth looking at how LEDs are hooked up to see why this is important. In
the old days, discrete LEDs were relatively power-hungry devices and the
recommended way to drive one reliably from a digital signal was as shown in
Figure \ref{figure:LED_drivers}a. The buffer/driver might be an inverter, as
shown here, or a transistor; the current-limiting resistor was to protect the
LED. Nowadays, GPIO pins have current-sink specifications within the range of
common LEDs, the supply voltage has gone down from \SI{5}{\volt} to
\SI{3.3}{\volt}, and so the resistor is often unnecessary. In fact, in tight
board layouts and some designs, when power traces are a scarce resource and
GPIOs are plentiful, the designer might choose to drive an LED as in Figure
\ref{figure:LED_drivers}b. Here, one GPIO is configured as a digital output
and driven HIGH; the other GPIO is configured as a digital output and driven
LOW. It even has the advantage, on the breadboard, that it doesn't matter which
way round the LED is hooked up. Most discrete LEDs are quite tolerant of
polarity reversal at \SI{3.3}{\volt}DC. It's not exactly recommended, but it
works. And there are lots of hardware designers in the world.\footnote{And some
of them are on drugs.}

External hardware pull-up or pull-down resistors, as used on the Intel 8255 and
MOS 6522 chips in the 1980s, are, of course, inaccessible to an adversary; CMOS
variants of the 8255 had internal pull-up resistors but they were not
programmable. (Automatically configured pull-up or pull-down resistors, which
are present in some devices, may be considered a special case of external
discrete hardware components.)

Programmable pull-up and pull-down resistors,
introduced in 1996 with the ATmega8 microcontroller \cite{Atmel2013} are far
more useful to an attacker because they can be set in nonsensical ways. The
ATmega series of microcontrollers have only pull-ups, but
the FreeBSD \texttt{gpio(3)} library contains pull-down methods as
well, for those chips that support them \cite{Microchip2018b,MediaTek2016}.

\begin{figure}[ht]
  \centering
  \includegraphics[width=0.6\columnwidth]{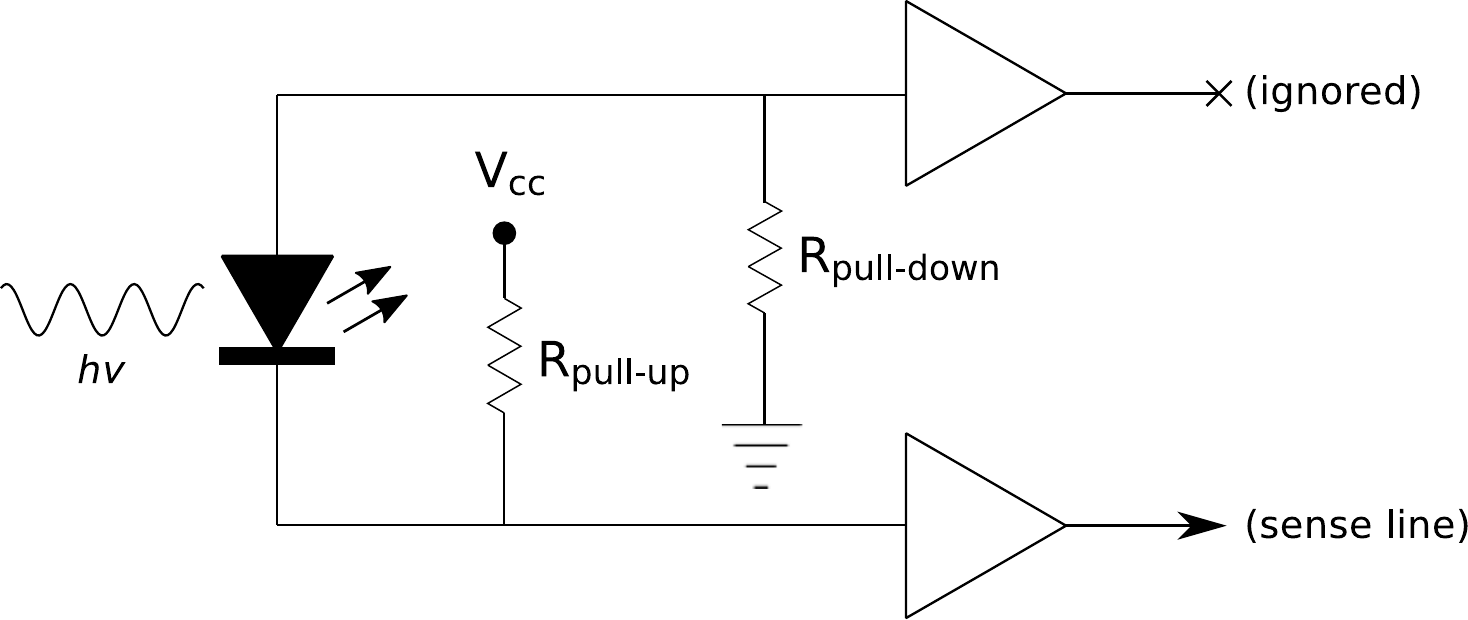}
  \caption{Programmable pull-up and pull-down resistors can be abused by
    setting them in non-sensical ways, like this, to reverse-bias an LED
    and sense it---as a photodiode---in the (preferred) photoconductive
    mode.}
  \label{figure:nonsensical}
\end{figure}

LEDs are known to work as photodiodes in both the conventional reverse-biased,
or photoconductive---the third quadrant of the $V/I$ plot---region, as well as
in the less commonly used forward-biased, or photovoltaic---quadrant IV---mode
\cite{Uiga1995}. The usual way a photodiode is hooked up is with a small
reverse bias, for improved speed and sensitivity. As shown in Figure
\ref{figure:conventional_photodiode}, the \SI{10}{\kilo\ohm} pull-up resistor
on the input port serves also to bias the photodiode with \SI{33}{\milli\volt};
the \SI{100}{\ohm} resistor protects the diode from a short circuit to ground
in bright light. Compare this circuit to Figure \ref{figure:nonsensical}, a
completely nonsensical---but legal---configuration of two GPIO ports and an
LED, identical in its physical wiring to the LED indicator driver in Figure
\ref{figure:LED_drivers}b right down to the polarity of the LED. The only
difference in Figure \ref{figure:nonsensical} is the direction of the GPIOs;
rather than configured as digital outputs, as the designer intended, they have
been re-configured by an attacker as digital inputs and the programmable
pull-up and pull-down resistors turned on. The topmost GPIO input is
ignored---its only purpose is to serve as a provider for a pull-down
resistor---and the bottom-most GPIO will see an approximately \SI{3.3}{\volt}
level from the ersatz photodiode in darkness, or \SI{0}{\volt} in the light.
That is enough, in testing, to sense logic levels from the LED and provide a
channel to the inside.

\newcommand{\anode}{\textcolor{red}{\LEFTcircle}}
\newcommand{\cathode}{\textcolor{blue}{\RIGHTcircle}}
\newcommand{\anomaly}{\textcolor{black}{\danger}}
\begin{table*}[ht]
  \begin{centering}
  \caption{Results of discrete LED reversing experiments. The response seen on
    both anode \emph{and} cathode for some LEDs is surprising. The response
    seen at the cathode---only---of the diffuse yellow LED is even more
    surprising; the initial hypothesis to explain apparent cathode emission
    (that photocurrents originating at the anode might be flowing through the
    diode to the cathode) was invalidated by the subsequent observation that
    one LED actually produced a photocurrent from the cathode terminal only.}
  \label{table:results_of_LED_reversing_experiments}
  \bigskip
  \begin{tabular}{lcccccccc}
    & \multicolumn{8}{c}{\large Excitation Wavelength} \\
    & \multicolumn{2}{c}{\SI{405}{\nano\metre} violet laser diode} &
      \multicolumn{2}{c}{\SI{532}{\nano\metre} green laser diode} &
      \multicolumn{2}{c}{\SI{640}{\nano\metre} red laser diode} &
      \multicolumn{2}{c}{white LED [not a laser]} \\
    Target Device & (anode) & (cathode)
      & (anode) & (cathode) & (anode) & (cathode) & (anode) & (cathode) \\
    \hline
    \SI{5}{\milli\metre} true green LED \rule{0pt}{3.5mm}
      & \anode & \cathode & \anode & \cathode
        & \anode & \cathode & \anode & \cathode \\
    \SI{5}{\milli\metre} pink LED & \anode \\
    \SI{5}{\milli\metre} white LED & \anode \\
    \SI{3}{\milli\metre} deep red LED\textsuperscript{\S} \\
    \SI{5}{\milli\metre} yellow diffuse LED
      & & \anomaly & & \anomaly & & \anomaly & & \anomaly \\
    \SI{5}{\milli\metre} blue LED
      & \anode & \cathode & \anode & \cathode
        & \anode & \cathode & \anode & \cathode \\
    \SI{5}{\milli\metre} red diffuse LED\textsuperscript{\S} \\
    \SI{5}{\milli\metre} UV LED
      & \anode & & \anode & & \anode & & \anode & \\
    \SI{5}{\milli\metre} green diffuse LED\textsuperscript{\S} \\
    \SI{5}{\milli\metre} ultrabright red LED\rule[-2mm]{0pt}{1ex}
      & \anode & \cathode & \anode & \cathode
        & \anode & \cathode & \anode & \cathode \\
    \hline
  \end{tabular}
  \end{centering}
  \rule{0pt}{0.2in}

  \vspace{2mm}
  Legend:
  \begin{description}
    \item[\anode] \hspace{-1em}Expected response
    \item[\cathode] \hspace{-1em}Unexpected response
    \item[\hspace{-0.15em}\anomaly] \hspace{-1em}Unexplainable response
  \end{description}

  \vspace{1mm}
  \textsuperscript{\S}\,No response from these devices
\end{table*}

\section{Threats and Countermeasures}\label{section:disk_drive_indicator}
What can be done with that vulnerability? After the publication of the 2002
journal article that started it all, I continued to search for further optical
emanations---especially from disk drive indicators---without success. I
remembered the first time I'd heard about LED photosensitivity and I wondered
if it could be exploited as a way in \cite{Mims1977a}. But the sticking point
was always how to do it with an unmodified target, as we had done with optical
TEMPEST. The vulnerability back then was in hardware: LED indicators on
numerous equipment simply leaked information.
That was what all the previous researchers---Wright, van Eck, Smulders, Kocher,
Kuhn, and Kubiak---had found \cite{vanEck1985, Wright1987, Smulders1990,
Kocher1999,
Kuhn2002, NSAtempest2007, Kubiak2017d}. All of these attacks exploited
unmodified, un-subverted---although inadequately shielded---systems. In
contrast, attacks like GUNMAN or Soft TEMPEST or VAGRANT require the attacker
to be able to install some kind of
malicious hardware or software on the system in order to cause the compromising
emanations to occur \cite{Maneki2007a,Kuhn1998a,NSA2013}.
\subsection{Covert Channels}
In the appendix of that 2002 paper, though, we described
another form of attack---an active attack---which required the attacker to
implant malicious code (malware) on the target system. This was a classical
covert channel ({\it vide supra}) in that it required two intentionally
communicating
processes---one inside the computer system and another outside. (This is the
difference between an accidental information leak and a covert channel.) It was
less
interesting, we thought, than the intrinsic vulnerability of an accidental
information leak, although much easier
to exploit, on the other hand, assuming the attacker had the capability to
emplace malicious hardware on the target system.

Now we have the Internet of Things (IoT) and in this world, malware is much
easier to introduce than before (because there are orders of magnitude more
devices in the world, hence manufacturers design in the capability to update
their software remotely [because otherwise it would be impossible to visit them
all] but hackers find out how to exploit those remote update features, and use
them to install malware on the devices). Consequently, many information
security researchers since have employed malware on the target system
to generate intelligible compromising optical emanations, greatly simplifying
the problem of receiving information (since the transmitted information can be
encoded sensibly, modulated onto a high-powered carrier, and sent under
relatively controlled conditions). The most prolific research group in the field
of malware-facilitated compromising electromagnetic emanations is Guri {\it et
al.}\ \cite{Sepetnitsky2014a,Guri2015a,Guri2016b,Guri2016c,Guri2017b,
Guri2017c, Guri2018a,Guri2018b,Guri2018d,Guri2018e,Guri2018f,
Guri2018g, Guri2018}.\footnote{Guri finally made the disk drive indicator
attack (\S\ref{section:disk_drive_indicator}) work in 2017 but he used a
malware to do it \cite{Guri2017a}.}
Interestingly, acoustical compromising emanations researchers have tended to
stick to the more difficult problem of intrinsic vulnerabilities
\cite{Asonov2004,Zhuang2005,Berger2006,Backes2010,Genkin2018a,Kubiak2018f}
although Guri is active there with his methodology as well
\cite{Guri2016a,Guri2016d,Mirsky2017,Guri2018c}.

Countermeasures to covert channels classically lie along a pair of conceptual
axes: prevention and detection. Covert channel analysis (CCA) is notoriously
time consuming and difficult, focusing on identification, characterisation, and
limitation of timing and storage remanences. CCA strives to block potential
covert channels by clearing shared areas of storage before they can be reused,
or by limiting the potential bandwidth of covert timing channels by adequately
synchronising access to shared resources. Neither of these mitigations can be
effective in the face of {\it post hoc} malware installed for the purpose of
receiving from a covert channel. Instead, prevention, detection, and removal of
malware (by means of continual system integrity checking, cryptographic signing
of software updates, and protected storage, for example) are needed.

\section{Experimental Results}
I found this vulnerability because I went looking for it. There was no
point to searching for vulnerabilities in a hardware configuration that does
not exist. That is, until Arduino came along (refer to the timeline in Table
\ref{table:timeline}). At
first, the logical approach seemed to be to look at the conventional ways that
LEDs are
hooked up to GPIO pins (Figure \ref{figure:LED_drivers}a), and the conventional
way that photodiodes are sensed (Figure \ref{figure:conventional_photodiode}),
and attempt to reconcile the two in such a way that available GPIOs might be
reconfigured to put an LED into reverse bias (Figure \ref{figure:nonsensical}).
But that never worked. It turns out that LEDs function best as photodiode
sensors in photovoltaic mode.

The results were startling (see Table
\ref{table:results_of_LED_reversing_experiments}). First of all, what was
thought to be the most promising approach---reverse-biasing the LED to sense it
like a conventional photodiode in photoconductive mode---did not work at all.
Secondly, 70 percent of the first batch of ten LEDs chosen at random from the
junk box (Figure \ref{figure:test_subjects}) exhibited strong photovoltaic
effects. Thirdly, photocurrent was commonly observed on both anode and cathode
leads. Finally, my hypothesis (that photocurrents originating at the anode were
flowing through the diode where they were also sensed at the cathode) was shot
down by the observation that one LED produced photocurrent on the cathode
terminal only. (Until the yellow diffuse LED is decapsulated and examined more
closely, it remains possible that the anode and cathode terminals of this one
device were misidentified.) No straightforward relationship amongst the
emission wavelength of an LED, the colour of the package (optical filter),
and the excitation wavelength to optical responsivity is apparent.

LEDs have a relatively wide emission spectrum---at least compared to
lasers---and respond to excitation wavelengths over an even wider spectral
range. Obviously, if the LED package incorporates an optical filter (such as
the familiar \SI{5}{\milli\metre} red diffuse LED in a red epoxy package) then
the filter will limit excitation by wavelengths outside the transmission
band of the filter.
\section{The Next Experiments}
If a directed energy optical attack is to be possible without a previous
system compromise, then it will be necessary to alter the CPU's internal state.
Already demonstrated is the capability to photovoltaically charge GPIO lines
above or below logic thresholds. What remains---in order to reject the null
hypothesis---is to show that physical effects might be propagated into the
interior.
\subsection{Threat Model, Countermeasures, and Limitations}
Ultimately, the goal is execution of arbitrary code. To get there, in stages,
we plan to attack the instruction fetch and decoding process first. Granted
that only one or a few bits of the instruction word, in the most optimistic
scenario, might turn out to be manipulable, the experimental design aims to
find, by exhaustive search, any window of opportunity that exists to set or
clear as much as a single bit in any instruction at any phase of the system
clock---including thoughtful exploration of the limits of allowable timing
exceptions.

Countermeasures (in hardware) are expected to include one-way buffering, tight
control of tolerances in setup-and-hold timing, cryptographically signed
instructions, as well as a CPU option to redirect immediately to a safe state
upon decoding any invalid instruction, up to a Hamming distance of 1, within
the current security context.\footnote{This might  be of interest in other high
security or high reliability realms such as cryptographic hardware or radiation
resistant processors.}

From the ATmega328P data sheet, it is clear that GPIO ports (buffered,
probably, by latches) are on the internal data bus \cite{Microchip2018b}. The
AVR core is a Harvard architecture, so we have no hope of directly modifying
instruction decoding from the data bus, but the programme counter is on the
data bus, and the stack resides in data SRAM which is also on the data bus. The
question is, can we impress a value on the data bus, through one of the I/O
modules, from a GPIO pin? That is what current experiments are trying to do. If
doable asynchronously, then an attacker could mess with ALU results, status and
control registers, or memory fetches. From experience, any effect at all is
likely stochastic \cite{Kim2014}. The most an attacker might be able to do is
to probabilistically flip one or more bits in the instruction stream between
decoding and execution; maybe the attacker could convert a valid instruction
into a slightly different one, or corrupt an instruction to an invalid opcode
that would be ignored (Figure \ref{figure:timing_diagram} shows the window of
opportunity).

Lacking specific internal implementation details---short of reverse engineering
the microcontroller from transistor doping regions and wires
\cite{Shirriff2017} in a microscopic die photo---we must experiment. To do it,
we need a fast, controllable optical modulator. Unfortunately, Pockels cells
tend to top out at about \SI{5}{\kilo\hertz}, so we are forced to use an
electro-optic (EO) modulator, which can be driven at more like
\SI{250}{\mega\hertz}.
\begin{figure}[ht]
  \centering
  \includegraphics[width=\columnwidth]{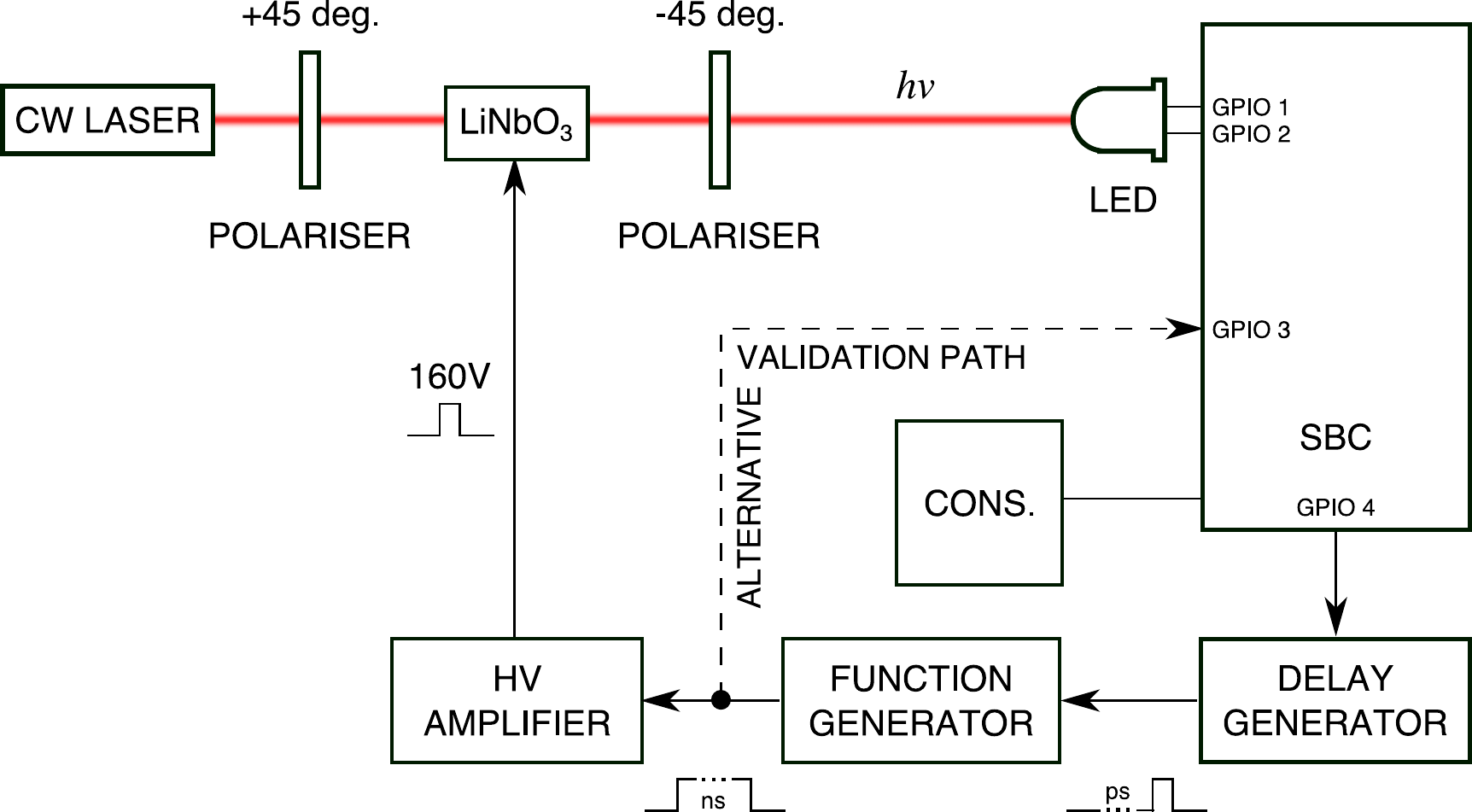}
  \caption{Phase II experimental setup (proposed).}
  \label{figure:EO_experiment}
\end{figure}
\subsection{Experimental Design for Phase II}
Figure \ref{figure:EO_experiment} shows the proposed experimental setup; a
visible continuous wave (CW) laser source---chosen for intensity and
directivity\footnote{Also useful later on for the remote attack.}---amplitude
modulated by a lithium niobate crystal between two crossed polarisers and aimed
at the LED under test, which is connected to two GPIO pins on the single board
computer (SBC) as in the previous assumptions about target hardware.
\begin{figure}[ht]
  \centering
  \includegraphics[width=\columnwidth]{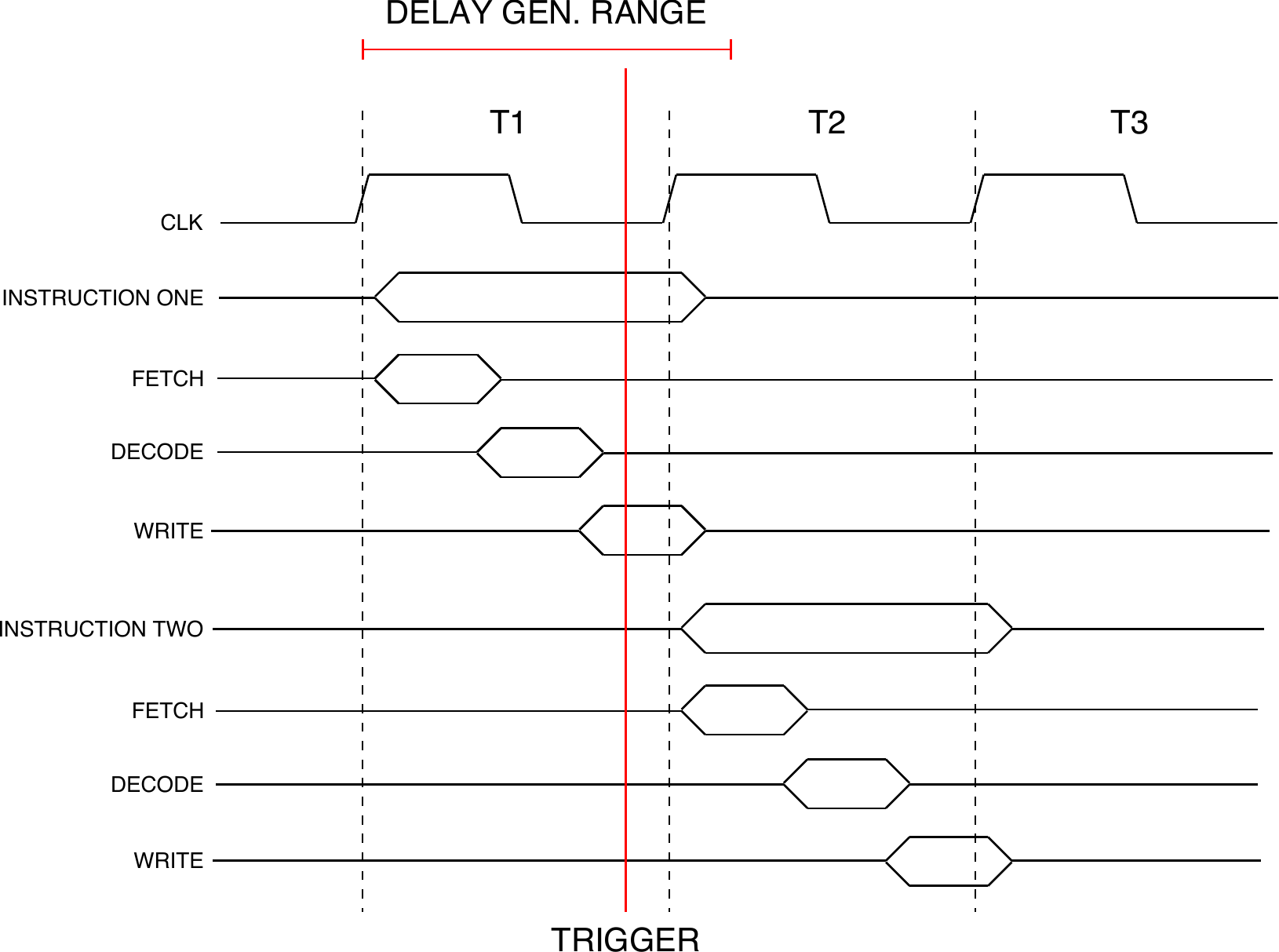}
  \caption{Notional timing diagram of a representative CPU. Only trivial
    pipelining is shown because it is irrelevant; the vulnerability, if it
    exists, must affect all execution units.}
  \label{figure:timing_diagram}
\end{figure}
The duty cycle of the laser illumination pulse is determined by a function
generator triggered by the SBC---via any other conveniently unused GPIO
pin---through a programmable delay generator. In essence, it's a lock-in
amplifier synchronised to a control signal emitted by the SBC; we use the delay
generator to hunt around before and after the synchronisation pulse, looking
for any vulnerable microsecond. Whilst the SBC is exercised by cycle-counted
assembly language code presenting each opcode in turn exactly once in a defined
time window, the delay generator will be swept across the width of the
microarchitecture's timing diagram, a coroutine watching concurrently for
unexpected results in any of the CPU registers (Figure
\ref{figure:timing_diagram}). Conceptually, the method is a hybrid of glitching
\cite{Skorobogatov2011a} and fuzzing \cite{vanSprundel2005}, except induced
through I/O pins instead of via the power supply.
\subsection{Validation Plan}
The expected limitations are severe: likely it may be possible only to force a
fixed value onto the internal data bus---for example, to force bits low but not
high---although even that much is enough to alter a running stream of
instructions; {\it e.g.}, to convert a branch instruction to some other kind,
thereby bypassing a security check in code.

\begin{figure}[ht]
  \centering
  \includegraphics[width=\columnwidth]{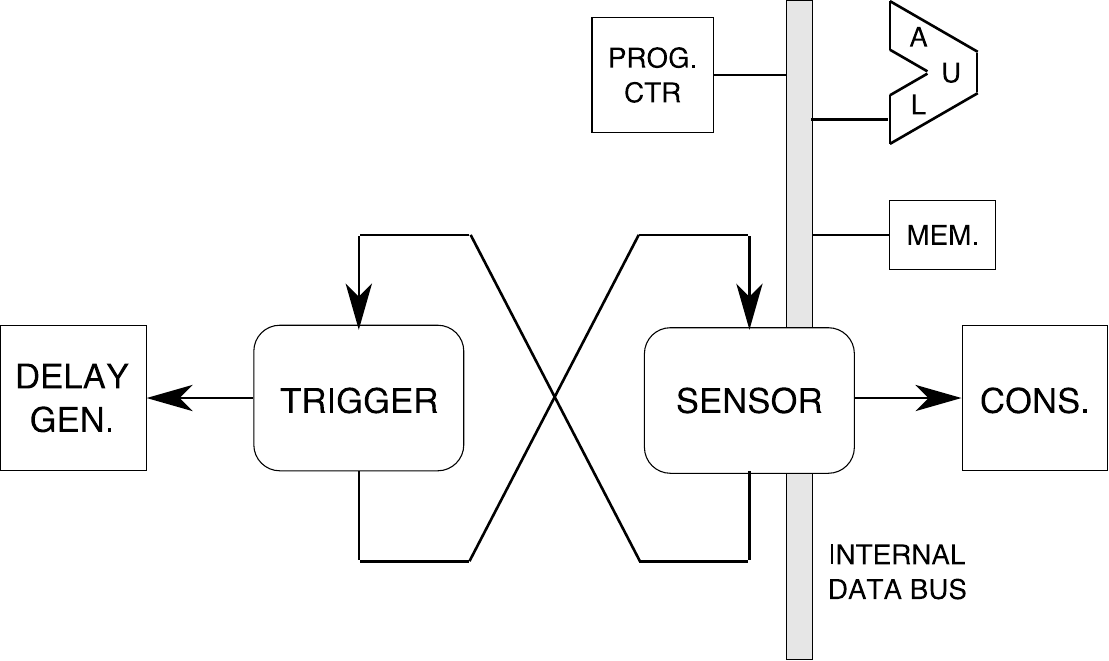}
  \caption{Coroutines on the CPU ensure a stable trigger and reliable detection
  of anomalous behaviour of anything on the internal data bus.}
  \label{figure:coroutines}
\end{figure}

The success criteria for this experiment are dependent on the detector (Figure
\ref{figure:coroutines}) and the injector (Figure \ref{figure:EO_experiment}).
To improve the chance of success, some independent variables are eliminated.
Firstly, we need only analyse one execution unit; the location of the trigger
within the fetch-execute cycle can be arbitrary, because we have a delay
generator. Secondly---at first---use the alternative validation path shown as a
dashed line in Figure \ref{figure:EO_experiment}. By applying logic levels
directly to the target pin, questions of optical intensity, aim, focus,
wavelength, and LED responsivity are moot. Once the vulnerable time window has
been found, then directed energy is brought back into the equation. In the
worst case, this method would validate the absence of race conditions or
setup/hold violations in the CPU support circuitry.

This kind of noisy, high-dimensional data analysis is exactly the kind of
scientific application for which machine learning (ML) is highly applicable
\cite{Asonov2004,Zhuang2005,Barisani2009a}. However, there are risks
\cite{Allen2019,Ghosh2019}. To mitigate the scientific hazard (primarily lack
of replication) of over-enthusiastic application of ML, we propose to use
different methods, beginning with SciKit-Learn
\URL{https://scikit-learn.org/stable/modules/outlier_detection.html} for
novelty and outlier detection \cite{INRIA2019}. The training phase presents a
difficulty; in contrast to typical use of ML, we don't {\it a priori} know what
stimulus will result in anomalous behaviour---we have to try all stimuli.
Turning the problem upside-down, therefore, we propose to train the neural
network (NN) on normal behaviour---instrumented by periodic snapshotting of all
registers including stack and ALU result---and then use the trained NN to
detect anomalous behaviour. For the second method, TensorFlow: use a support
vector machine (SVM) unsupervised to learn a soft boundary from the normal
training data \cite{Abadi2016}.
Funding to purchase the necessary apparatus to begin Phase II experiments is in
the proposal stage. Finally, in Phase III, plan to attack the in-built LED of a
standard SBC, to prove that it can be done in practice on hardware in the real
world.
\section{Summary and Conclusions}
Possible right now is for malware to freely repurpose LED status indicators
into optical receivers capable of ingesting information at least in the high
\si{\kilo\bit\per\second} range. This has been demonstrated in the lab. The
chance that an attacker will find an agreeable LED available for use seems to
be 70 percent, although that number depends on the hardware designer making a
specific kind of mistake.

These experiments are the tip of the iceberg regarding methods for
sensing---GPIO digital bipolar, GPIO digital single-ended, cathode emission,
analogue GPIO (single or double-ended)---and improvements to energy transfer for
Rowhammer-style attacks are all waiting to be explored.
\section*{Acknowledgements}
I want to thank the anonymous reviewers for their careful reading and
thoughtful suggestions.
Thanks to Ireneusz Kubiak in the EMC Laboratory of Wojskowego Instytutu
\L{}\c{a}czno\'{s}ci, Zegrze, Poland; Chris Roberts in Denver, Colorado;
Jos\'{e} Lopes Esteves of ANSSI, Henry Gold of API Technologies Corp., and Ivo
Kauffmann of CN Rood in the Netherlands for critical evaluation and feedback on
the idea. Portions of this material were workshopped in and after an oral
presentation at EMC Europe 2018 in Amsterdam \cite{Loughry2018a}. The quotation
in the title comes from a Daffy Duck cartoon \cite{Jones1953a}.
\IEEEtriggeratref{48}
\bibliographystyle{IEEEtran}

\begin{thebibliography}{10}
\providecommand{\url}[1]{#1}
\csname url@samestyle\endcsname
\providecommand{\newblock}{\relax}
\providecommand{\bibinfo}[2]{#2}
\providecommand{\BIBentrySTDinterwordspacing}{\spaceskip=0pt\relax}
\providecommand{\BIBentryALTinterwordstretchfactor}{4}
\providecommand{\BIBentryALTinterwordspacing}{\spaceskip=\fontdimen2\font plus
\BIBentryALTinterwordstretchfactor\fontdimen3\font minus
  \fontdimen4\font\relax}
\providecommand{\BIBforeignlanguage}[2]{{%
\expandafter\ifx\csname l@#1\endcsname\relax
\typeout{** WARNING: IEEEtran.bst: No hyphenation pattern has been}%
\typeout{** loaded for the language `#1'. Using the pattern for}%
\typeout{** the default language instead.}%
\else
\language=\csname l@#1\endcsname
\fi
#2}}
\providecommand{\BIBdecl}{\relax}
\BIBdecl

\bibitem{Mims1973b}
F.~M. Mims~III, \emph{Light Emitting Diodes}.\hskip 1em plus 0.5em minus
  0.4em\relax Indianapolis, Indiana, USA: Howard W.~Sams \& Co., Inc., 1973.

\bibitem{Mims1977a}
F.~M. Mimms, ``Using {LED}'s as light detectors,'' \emph{Popular Electronics},
  vol.~11, no.~5, pp. 86--88, May 1977.

\bibitem{Mims1979a}
F.~M. Mims~III, ``Bidirectional optoisolator puts two {LED}s nose to nose,''
  \emph{Electronics}, vol.~52, no.~10, p. 127, 10th May 1979.

\bibitem{MERL2003a}
P.~Dietz, W.~Yerazunis, and D.~Leigh, ``Very low-cost sensing and communication
  using bidirectional {LED}s,'' Mitsubishi Electronics Research Laboratories
  (MERL), 201 Broadway, Cambridge, Massachusetts 02139, USA, Tech. Rep.
  TR2003-35, July 2003.

\bibitem{Karadaglic2007a}
R.~Stojanovi\'{c} and D.~Karadagli\'{c}, ``Single {LED} takes on both
  light-emitting and detecting duties,'' \emph{Electronic Design}, vol.~55,
  no.~16, pp. 53--54, 18th July 2007, uRL:
  \url{https://www.electronicdesign.com/lighting/single-led-takes-both-light-emitting-and-detecting-duties}.

\bibitem{Mims2014a}
F.~M. Mims~III, ``How to use {LED}s to detect light,'' \emph{Make}, vol.~36, p.
  136, 8th January 2014.

\bibitem{Thornton2019}
S.~Thornton, ``Understanding delay for {I/O}: Using {Arduino} functions vs.\
  coding the {MCU},'' EEworld Online
  \URL{https://www.eeworldonline.com/understanding-delay-for-i-o-using-arduino-functions-vs-coding-the-mcu/},
  4 February 2019.

\bibitem{Stringfellow1997}
G.~B. Stringfellow and M.~G. Craford, \emph{High Brightness Light Emitting
  Diodes}, ser. Semiconductors and Semimetals.\hskip 1em plus 0.5em minus
  0.4em\relax San Diego, California: Academic Press, 1997, vol.~48.

\bibitem{Loughry2002a}
J.~Loughry and D.~A. Umphress, ``Information leakage from optical emanations,''
  \emph{{ACM} Transactions on Information and System Security}, vol.~5, no.~3,
  pp. 262--289, August 2002.

\bibitem{Lampson1973}
B.~W. Lampson, ``A note on the confinement problem,'' \emph{Comm.\ ACM},
  vol.~16, no.~10, pp. 613--615, October 1973.

\bibitem{Domburg2006}
J.~Domburg, ``Optical mouse cam,'' \url{http://spritesmods.com/?art=mouseeye},
  2006.

\bibitem{Rieback2006}
M.~R. Rieback, B.~Crispo, and A.~S. Tanenbaum, ``Is your cat infected with a
  computer virus?'' in \emph{Proceedings of the Fourth Annual {IEEE}
  International Conference on Pervasive Computing and Communications ({PERCOM
  '06})}.\hskip 1em plus 0.5em minus 0.4em\relax Washington, DC, USA: IEEE
  Computer Society, 2006, pp. 169--179.

\bibitem{Nassi2017a}
B.~Nassi, A.~Shamir, and Y.~Elovici, ``Oops!\ldots {I} think {I} scanned a
  malware,'' arXiv:1703.07751v1 [cs.CR], 22nd March 2017.

\bibitem{Jennings1990}
K.~Jennings, \emph{The Devouring Fungus: Tales of the Computer Age}.\hskip 1em
  plus 0.5em minus 0.4em\relax W.~W.~Norton and Company, Inc., 1990.

\bibitem{Rosenheim1983}
D.~E. Rosenheim, ``Installation of the first production 701,'' \emph{Annals of
  the History of Computing}, vol.~5, no.~2, pp. 146--147, April--June 1983.

\bibitem{ONsemi2018}
{ON Semiconductor}, ``Wafer-level chip-scale package ({WLCSP}) at {ON
  Semiconductor},'' Fairchild Semiconductor, Application Note 5075/D, October
  2018.

\bibitem{Micron1984}
{Micron Technology}, \emph{{IS32 Optic Ram data sheet}}, Boise, Idaho, USA, May
  1984.

\bibitem{Whitehead1984}
D.~G. Whitehead, I.~Mitchell, and P.~V. Mellor, ``A low-resolution vision
  sensor,'' \emph{J.\ Phys.\ E: Sci.\ Instrum.}, vol.~17, pp. 653--656, 1984.

\bibitem{Kurz1994}
H.~Kurz, ``Museum f\"{u}r mi{\ss}brauchte bauteile,'' Kurzschluss
  \URL{http://www.kurzschluss.com/kuckuck/kuckuck.html}, 1994.

\bibitem{Upton2015}
L.~Upton, ``Xenon death flash: a free physics lesson,'' Raspberry Pi blog
  \URL{https://www.raspberrypi.org/blog/xenon-death-flash-a-free-physics-lesson/},
  9 February 2015.

\bibitem{Atmel2013}
{Atmel Corporation}, ``{ATmega8} data sheet,'' 2013.

\bibitem{Microchip2018b}
{Microchip Technology, Inc.}, ``{ATmega48A/PA/88A/PA/168A/PA/328/P
  megaAVR\textsuperscript{\textregistered}} data sheet,'' 2018.

\bibitem{MediaTek2016}
{MediaTek, Inc.}, ``{MediaTek MT7688} datasheet,'' 15th April 2016.

\bibitem{Uiga1995}
E.~Uiga, \emph{Optoelectronics}.\hskip 1em plus 0.5em minus 0.4em\relax
  Englewood Cliffs, New Jersey: Prentice-Hall, 1985.

\bibitem{vanEck1985}
W.~{van Eck}, ``Electromagnetic radiation from video display units: An
  eavesdropping risk?'' \emph{Computer Security}, vol.~4, no.~4, pp. 269--286,
  December 1985.

\bibitem{Wright1987}
P.~Wright, \emph{Spycatcher: The Candid Autobiography of a Senior Intelligence
  Officer}.\hskip 1em plus 0.5em minus 0.4em\relax New York: Viking Press,
  1987.

\bibitem{Smulders1990}
P.~Smulders, ``The threat of information theft by reception of electromagnetic
  radiation from {RS-232} cables,'' \emph{Computer Security}, vol.~9, no.~1,
  pp. 53--58, February 1990.

\bibitem{Kocher1999}
P.~Kocher, J.~Jaffe, and B.~Jun, ``Differential power analysis,'' in
  \emph{Proceedings of the 19th Annual International Cryptology Conference on
  Advances in Cryptology {CRYPTO 99}}, Santa Barbara, California, 15--19 August
  1999, pp. 388--397.

\bibitem{Kuhn2002}
M.~G. Kuhn, ``Optical time-domain eavesdropping risks of {CRT} displays,'' in
  \emph{Proceedings of the 2002 {IEEE} Symposium on Security and
  Privacy}.\hskip 1em plus 0.5em minus 0.4em\relax Berkeley, California: {IEEE}
  Computer Society, 12--15 May 2002, pp. 3--18.

\bibitem{NSAtempest2007}
{National Security Agency}, ``{TEMPEST}: A signal problem,'' \emph{Cryptologic
  Spectrum}, 1972.

\bibitem{Kubiak2017d}
I.~Kubiak, ``{LED} printers and safe fonts as an effective protection against
  the formation of unwanted emission,'' \emph{Turkish Journal of Electrical
  Engineering and Computer Sciences}, vol.~25, no.~5, pp. 4268--4279, 2017.

\bibitem{Maneki2007a}
S.~Maneki, ``Learning from the enemy: the {GUNMAN} project,'' \emph{United
  States Cryptologic History {Series VI}}, vol.~13, 8th January 2007.

\bibitem{Kuhn1998a}
M.~G. Kuhn and R.~J. Anderson, ``Soft {Tempest}: Hidden data transmission using
  electromagnetic emanations,'' in \emph{Second International Workshop on
  Information Hiding ({IH'98})}, Portland, Oregon, USA, 15--17 April 1998, pp.
  124--142.

\bibitem{NSA2013}
{Advanced Network Technology (ANT) Division}, ``{RANGEMASTER},'' National
  Security Agency, Fort Meade, Maryland, USA, ANT Product Data, 24 July 2008.

\bibitem{Sepetnitsky2014a}
V.~Sepetnitsky, M.~Guri, and Y.~Elovici, ``Exfiltration of information from
  air-gapped machines using monitor's {LED} indicator,'' in \emph{2014 IEEE
  Joint Intelligence and Security Informatics Conference}, 2014, pp. 264--7.

\bibitem{Guri2015a}
M.~Guri, M.~Monitz, Y.~Mirski, and Y.~Elovici, ``{BitWhisper}: Covert signaling
  channel between air-gapped computers using thermal manipulations,''
  arXiv:1503.07919, 26th March 2015.

\bibitem{Guri2016b}
M.~Guri, O.~Hasson, G.~Kedma, and Y.~Elovici, ``{VisiSploit}: An optical
  covert-channel to leak data through an air-gap,''
  \url{https://arxiv.org/abs/1607.03946}, 13 July 2016.

\bibitem{Guri2016c}
M.~Guri, M.~Monitz, and Y.~Elovici, ``{USBee}: Air-gap covert-channel via
  electromagnetic emission from {USB},'' arXiv preprint \url{arXiv:1608.08397}
  [cs.CR], 30th August 2016.

\bibitem{Guri2017b}
M.~Guri, B.~Zadov, A.~Daidakulov, and Y.~Elovici, ``{xLED}: Covert data
  exfiltration from air-gapped networks via router {LEDs},'' arXiv preprint
  1706.01140 [cs.CR], 4th June 2017.

\bibitem{Guri2017c}
M.~Guri, D.~Bykhovsky, and Y.~Elovici, ``{aIR-Jumper}: Covert air-gap
  exfiltration/infiltration via security cameras \& infrared ({IR}),'' arXiv
  preprint arXiv:1709.05742 [cs.CR], 18th September 2017.

\bibitem{Guri2018a}
M.~Guri, B.~Zadov, A.~Daidakulov, and Y.~Elovici, ``{ODINI}: Escaping sensitive
  data from {Faraday}-caged, air-gapped computers via magnetic fields,'' arXiv
  preprint 1802.02700 [cs.CR], 8th February 2018.

\bibitem{Guri2018b}
M.~Guri, A.~Daidakulov, and Y.~Elovici, ``{MAGNETO}: Covert channel between
  air-gapped systems and nearby smartphones via {CPU}-generated magnetic
  fields,'' Ben-Gurion University of the Negev, Cyber Security Research Center,
  arXiv preprint 1802.02317 [cs.CR], 7th February 2018.

\bibitem{Guri2018d}
M.~Guri and Y.~Elovici, ``Bridgeware: the air-gap malware,'' \emph{Comm.\ ACM},
  vol.~61, no.~4, pp. 74--82, April 2018.

\bibitem{Guri2018e}
M.~Guri, B.~Zadov, D.~Bykhovsky, and Y.~Elovici, ``{PowerHammer}: Exfiltrating
  data from air-gapped computers through power lines,'' arXiv:1804.04014
  [cs.CR], 10th April 2018.

\bibitem{Guri2018f}
M.~Guri, ``{BeatCoin}: Leaking private keys from air-gapped cryptocurrency
  wallets,'' arXiv preprint 1804.08714, 23 April 2018.

\bibitem{Guri2018g}
M.~Guri, B.~Zadov, A.~Daidakulov, and Y.~Elovici, ``Covert data exfiltration
  from air-gapped networks via switch and router {LED}s,'' in \emph{16th Annual
  Conference on Privacy, Security and Trust ({PST})}, Belfast, UK, 28--30
  August 2018, pp. 269--274.

\bibitem{Guri2018}
M.~Guri and M.~Monitz, ``{LCD TEMPEST} air-gap attack reloaded,'' in \emph{2018
  {IEEE} International Conference on the Science of Electrical Engineering
  ({ICSEE})}, Eilat, Israel, 12--14 December 2018.

\bibitem{Guri2017a}
M.~Guri, B.~Zadov, E.~Arias, and Y.~Elovici, ``{LED-it-GO}: Leaking (a lot of)
  data from air-gapped computers via the (small) hard drive {LED},''
  \url{http://cyber.bgu.ac.il/advanced-cyber/system/files/LED-it-GO_0.pdf},
  2017.

\bibitem{Asonov2004}
D.~Asonov and R.~Agrawal, ``Keyboard acoustic emanations,'' in
  \emph{Proceedings of the 2004 {IEEE} Symposium on Security and Privacy},
  Oakland, California, May 9--12, 2004, pp. 3--11.

\bibitem{Zhuang2005}
L.~Zhuang, F.~Zhou, and J.~Tygar, ``Keyboard acoustic emanations revisited,''
  in \emph{Proceedings of the 12th {ACM} Conference on Computer and
  Communications Security}, Alexandria, Virginia, November 7--11, 2005, pp.
  373--382.

\bibitem{Berger2006}
Y.~Berger, A.~Wool, and A.~Yeredor, ``Dictionary attacks using keyboard
  acoustic emanations,'' in \emph{CCS'06}, Alexandria, Virginia, October
  30--November 3, 2006, pp. 245--254.

\bibitem{Backes2010}
M.~Backes, M.~D\"{u}rmuth, S.~Gerling, M.~Pinkal, and C.~Sporleder, ``Acoustic
  side-channel attacks on printers,'' in \emph{Proceedings of the 19th {USENIX}
  Security Symposium}, Washington, D.C., August 11--13, 2010.

\bibitem{Genkin2018a}
D.~Genkin, M.~Pattani, R.~Schuster, and E.~Tromer, ``Synesthesia: Detecting
  screen content via remote acoustic side channels,'' arXiv:1809.02629 [cs.CR],
  7th September 2018.

\bibitem{Kubiak2018f}
I.~Kubiak, ``The typewriters as the sources of the sensitive acoustic
  signals,'' in \emph{25th International Congress on Sound and Vibration
  ({ICSV25})}, Hiroshima, Japan, 8--12 July 2018.

\bibitem{Guri2016a}
M.~Guri, Y.~Solewicz, A.~Daidakulov, and Y.~Elovici, ``Fansmitter: Acoustic
  data exfiltration from (speakerless) air-gapped computers,'' arXiv ePrint
  1606.05915, 2016.

\bibitem{Guri2016d}
------, ``{SPEAKE(a)R}: Turn speakers to microphones for fun and profit,''
  \URL{http://cyber.bgu.ac.il/advanced-cyber/system/files/SPEAKE(a)R.pdf},
  2016.

\bibitem{Mirsky2017}
Y.~Mirsky, M.~Guri, and Y.~Elovici, ``{HVACKer}: Bridging the air-gap by
  manipulating the environment temperature,'' \emph{Magdeburger Journal zur
  Sicherheitsforschung}, vol.~14, pp. 815--829, September 2017.

\bibitem{Guri2018c}
M.~Guri, Y.~Solwicz, A.~Daidakulov, and Y.~Elovici, ``{MOSQUITO}: Covert
  ultrasonic transmissions between two air-gapped computers using
  speaker-to-speaker communication,'' arXiv preprint 1803.03422v1 [cs.CR], 9th
  March 2018.

\bibitem{Kim2014}
Y.~Kim, R.~Daly, J.~Kim, C.~Fallin, J.~H. Lee, D.~Lee, C.~Wilkerson, K.~Lai,
  and O.~Mutlu, ``Flipping bits in memory without accessing them: An
  experimental study of {DRAM} disturbance errors,'' in \emph{41st
  International Symposium on Computer Architecture ({ISCA'14}}, Minneapolis,
  Minnesota, USA, June 14--18, 2014.

\bibitem{Shirriff2017}
K.~Shirriff, ``Inside the 74181 {ALU} chip: die photos and reverse
  engineering,''
  \url{http://www.righto.com/2017/01/die-photos-and-reverse-engineering.html},
  January 2017.

\bibitem{Skorobogatov2011a}
S.~Skorobogatov, ``Fault attacks on secure chips: from glitch to flash,'' in
  \emph{Design and Security of Cryptographic Algorithms and Devices ({ECRYPT
  II})}, Albena, Bulgaria, 29 May--3rd June 2011.

\bibitem{vanSprundel2005}
I.~van Sprundel, ``Fuzzing,'' in \emph{22nd Chaos Communication Congress
  ({22C3})}, Berlin, Germany, 27--30 December 2005.

\bibitem{Barisani2009a}
A.~Barisani and D.~Bianco, ``Sniffing keystrokes with lasers/voltmeters: Side
  channel attacks using optical sampling of mechanical energy and power line
  leakage,'' in \emph{{CanSecWest} 2009}, Vancouver, British Columbia, Canada,
  March 16--20 2009.

\bibitem{Allen2019}
G.~Allen, ``Machine learning and statistics: Applications in genomics and
  computer vision,'' in \emph{{AAAS} Annual Meeting}, Washington, DC, 15
  February 2019.

\bibitem{Ghosh2019}
P.~Ghosh, ``{AAAS}: Machine learning `causing science crisis','' BBC News
  \URL{https://www.bbc.com/news/science-environment-47267081}, 16 February
  2019.

\bibitem{INRIA2019}
\emph{{scikit-learn}: Machine Learning in {Python}}, INRIA, France, 2019.

\bibitem{Abadi2016}
M.~Abadi, P.~Barham, J.~Chen, Z.~Chen, A.~Davis, J.~Dean, M.~Devin,
  S.~Ghemawat, G.~Irving, M.~Isard, M.~Kudlur, J.~Levenberg, R.~Monga,
  S.~Moore, D.~G. Murray, B.~Steiner, P.~Tucker, V.~Vasudevan, P.~Warden,
  M.~Wicke, Y.~Yu, and X.~Zheng, ``{TensorFlow}: A system for large-scale
  machine learning,'' in \emph{Proceedings of the 12th {USENIX} Symposium on
  Operating Systems Design and Implementation ({OSDI} '16)}, Savannah, Georgia,
  USA, 2--4 November 2016.

\bibitem{Loughry2018a}
J.~Loughry, ``Optical {TEMPEST},'' in \emph{International Symposium and
  Exhibition on Electromagnetic Compatibility ({EMC Europe} 2018)}, Amsterdam,
  The Netherlands, 27--30 August 2018.

\bibitem{Jones1953a}
C.~M. Jones, \emph{Duck Dodgers in the $24\frac{1}{2}$th Century}, ser. Merrie
  Melodies.\hskip 1em plus 0.5em minus 0.4em\relax Warner Bros., 1953.

\end{thebibliography}

\end{document}